\documentclass[10pt,twocolumn,prl,aps,floatfix,superscriptaddress,longbibliography]{revtex4-2}

\usepackage{color}
\usepackage{graphicx}
\usepackage{physics}
\usepackage{amsthm}
\usepackage{amsmath}
\usepackage{amssymb}
\usepackage{enumerate}
\usepackage{placeins}
\usepackage{booktabs}
\usepackage{dsfont}
\usepackage{yfonts}
\usepackage{hyperref}

\newcommand{\Eqref}[1]{Eq.~\eqref{#1}}

\usepackage[color=green!60]{todonotes}

\AtBeginDocument{%
\heavyrulewidth=.08em
\lightrulewidth=.05em
\cmidrulewidth=.03em
\belowrulesep=.65ex
\belowbottomsep=0pt
\aboverulesep=.4ex
\abovetopsep=0pt
\cmidrulesep=\doublerulesep
\cmidrulekern=.5em
\defaultaddspace=.5em
}

\newif\ifarXiv
\arXivtrue 
\ifarXiv

\usepackage{pdfpages} % include pdfs
\usepackage{pgffor} % for loops
\usepackage{xr} % referencing between files

\makeatletter
\AtBeginDocument{\let\LS@rot\@undefined}
\makeatother

\def\supplementfilename{quench_supplement}

\pdfximage{\supplementfilename.pdf}
\def\numbersupplementpages{\the\pdflastximagepages}

\fi % end ifarXiv

\begin{document}

\title{Measuring post-quench entanglement entropy through density correlations}

\author{Adrian Del Maestro}
\affiliation{Department of Physics and Astronomy, University of Tennessee, Knoxville, TN 37996, USA}
\affiliation{Min H. Kao Department of Electrical Engineering and Computer Science, University of Tennessee, Knoxville, TN 37996, USA}

\author{Hatem Barghathi}
\affiliation{Department of Physics and Astronomy, University of Tennessee, Knoxville, TN 37996, USA}

\author{Bernd Rosenow}
\affiliation{Institut f\"ur Theoretische Physik, Universit\"at Leipzig, D-04103, Leipzig, Germany} 

\begin{abstract}
Following  a sudden change of interactions in an integrable system of one-dimensional fermions, we analyze the dependence of the static structure factor on the observation time after the quantum quench. At small waiting times after the quench, we map the system to non-interacting bosons such that we are able to extract their occupation numbers from the Fourier transform of the density-density correlation function, and use these to compute a bosonic entropy from a diagonal ensemble. By comparing this bosonic entropy with the asymptotic steady state entanglement entropy per fermion computed with exact diagonalization we find excellent agreement.  
\end{abstract}

\maketitle

Recent experimental and theoretical advances have elucidated the time evolution of spatial entanglement for a system that remains in a pure quantum state after a sudden change in the Hamiltonian.  
In the steady state asymptotic limit, the entanglement entropy becomes extensive in the size of a spatial subsystem \cite{Calabrese:2005oi,Kaufman:2016ep}, and plays the role of a thermodynamic entropy arising from a microcanonical ensemble with the same energy density \cite{Srednicki:1994xg,Rigol:2007md,Rigol:2008sl,Polkovnikov:2011gg}.

Here we show that the entanglement entropy density after an interaction quench can be obtained from the diagonal ensemble density matrix \cite{Polkovnikov:2011gg,Santos:2011pc,Dora:2012,Gurarie:2013mu,Piroli:2017un}  of non-interacting bosons, whose properties can be determined from density-density correlations. Such correlations are readily accessible via current experimental technologies in a quantum gas via Bragg spectroscopy \cite{Altman:2004aa,Kuhnle:2010os,Boll:2016oa,Yang:2018po} without the need to directly measure coherences.
This result presents a complementary route to the experimental measurement of entanglement entropy in quantum systems that is not based on the creation of replicas \cite{Calabrese:2004ll, Daley:2012bd,Islam:2015cm,Kaufman:2016ep}, and highlights the 
role played by entanglement dynamics in generating an effective thermodynamic description that underlies our current framework of quantum statistical mechanics.

As the study of quantum quenches in one-dimensional systems has been very fruitful in understanding thermalization in closed quantum systems \cite{Cazalilla:2006bb,Lauchli:2008wb,Sabio:2010sk, Dora:2012,Ilievski:2015ap,Alba:2015et,Essler:2016sc,Alba:2017ph, Jansen:2019et}, we here consider an integrable model of interacting spinless fermions in one spatial dimension and obtain an explicit formula for the time dependence of the post-quench static structure factor via bosonization. The occupation numbers of bosonic modes can then be used to compute the entropy within the framework of an effective diagonal ensemble. This entropy is compared to the extensive asymptotic spatial entanglement entropy found via large scale exact diagonalization of the underlying fermions and we find excellent agreement between the two.  Our results can be readily adapted to an experimental protocol for measuring the entanglement after a quantum quench in trapped one dimensional quantum gases. 

\emph{Fermions with finite range interactions:} In equilibrium, all thermodynamic quantities characterizing interacting fermions in one dimension can be computed via a mapping to a thermal ensemble of non-interacting bosons \cite{delft98,giamarchi:2004qu}.
In particular, the thermal bosonic entropy can be directly  computed from knowledge of  the average bosonic mode occupancy. In the following, we describe how this concept can be generalized to a non-equilibrium situation after a quantum quench, and how the 
bosonic mode occupancies can be determined experimentally.  The starting point is an analysis of the Fourier transform of the density-density correlation function, known as the static structure factor. 

We consider a system of $N$ fermions on a one-dimensional lattice with $L$ sites, described by a Hamiltonian $H_0 + \theta(t) H_1$, which undergo a quantum quench at time $t=0$. The fermions are described by  creation and annihilation  operators  $c_i^\dagger$, $c_i$ with commutation relations $\{c_i, c_j^\dagger \} = \delta_{ij}$.  One can then define the  density operator $\rho_i = c_i^\dagger c_i$, with an average density  $\rho_0 = N/L$. We now consider the density operator $\rho_i(t)$ in the Heisenberg picture, evaluated at  time $t$ after the quantum quench, and define the density-density correlation function 
%
%**************************************************************************************
\begin{equation}
g_2(i-j; t) = \frac{\langle {\rho}_i(t)  {\rho}_j(t)  \rangle}{\rho_0^2} - \frac{\delta_{i,j}}{\rho_0}   \ \ .
\end{equation}
%*************************************************************************
%
We stress that $g_2$ is an equal time correlation function, with $t$ denoting the observation time after the quantum quench. Via a Fourier transform we obtain the static structure factor $s(q;t)$ after observation time $t$ as
%
%**************************************************************
\begin{equation}
s(q;t) = 1 + \rho_0 \sum_{r=0}^{L/2-1} \left[g_2(r;t)-1 \right] \mathrm{e}^{-i q r}\ .
\label{eq:Sqtdef}
\end{equation}
%*************************************************************************
%

\begin{figure}[t]
\begin{center}
\includegraphics[width=\columnwidth]{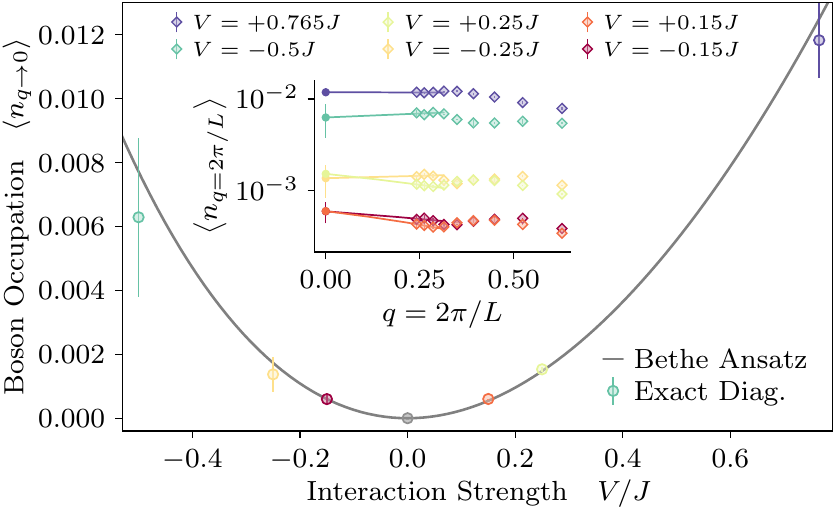}
\end{center}
\caption{{Bosonic density mode occupations.} The expectation value $\expval{n_q}$ for $q\to 0$ obtained via finite size scaling (inset) is compared with the prediction from Luttinger liquid theory, and we find very good agreement over a wide range of interaction strengths. Symbol colors are used to indicate different interaction strengths.}
\label{fig:nLL}
\end{figure}

Bosonization methods can be used to obtain a prediction for the observation time dependence of the structure factor, which in turn allows one to extract the occupation numbers of bosonic modes after the quantum quench. The result of such a calculation for a Luttinger liquid (LL) with quadratic Hamiltonian (shown in the supplementary material \cite{supplement}) is 
%
%*************************  time dependent LL structure factor ********************
\begin{equation}
s_{\rm LL}(q; t) = s_{LL}(q)  \left[ 2 \expval{n_q} + 1 + \sinh(2 \beta_q) \cos(2 \omega_q t)\right]\ ,
\label{eq:SqtSM}
\end{equation}
%*********************************************************************
%
where $\beta_q$ parametrizes the transformation between free fermions and the eigenstates of the post-quench Hamiltonian, and $\omega_q$ is the 
dispersion relation of density waves after the quantum quench. Above,  $s_{LL}(q)$ is the equilibrium ground state static structure factor, which at low energies and long wavelengths, is given by \cite{giamarchi:2004qu}
\begin{equation}
    s_{LL}(q) \underset{q\to0}{=}\frac{K |q|}{2k_{\rm F}} \ .
\label{eq:LLsq}
\end{equation}
The Luttinger parameter $K$ is related to the strength of interactions and $k_{\rm F}$ is the Fermi wavevector.  Considering the ratio $s_{\rm LL}(q; t)/s_{LL}(q)$ allows us to extract the bosonic occupation $\expval{n_q}$ from the time-independent part of this expression.  For a LL with a point-like (i.e.\@ momentum independent) interaction, one finds that $\expval{n_q} \equiv \expval{n_{q=0}} = \tfrac{1}{4} \qty(K+\tfrac{1}{K}-2)$  independent of $q$. For more realistic Hamiltonians, the interaction is momentum dependent, giving rise to a $q$-dependent bosonic mode occupancy $\expval{n_q}$ after the quench.  In addition, band curvature effects lead to a coupling between bosonic modes \cite{Haldane:1981bk,Samokhin:1998mc,Pereira:2006ds,DelMaestro:2010ib}, such that the mode occupancy $\expval{n_q}$ is no longer a constant of motion but acquires a time dependence. We expect these non-linear effects to be suppressed by at least a factor of $\expval{n_q}$, which is small (see Fig.~\ref{fig:nLL}).  In our numerical analysis, we use Eq.~(\ref{eq:SqtSM}) only for short times $t$ after the quench, for which the mode coupling has not yet had any appreciable effect.

Staying within the Luttinger liquid model, the Hamiltonian is quadratic in boson operators and thus the entropy of the density wave excitations can be computed exactly.  In this picture, the bosonic mode occupation operator $n_q$ commutes with the post-quench Hamiltonian, and thus all its higher order correlations are time-independent conserved quantities. When computing them using a squeezing transformation \cite{Cazalilla:2006bb,Iucci:2009ca,Dora:2011cf} between the original and  post-quench bosons \cite{supplement}, one finds for instance $\langle n_q^2 \rangle = 2 \langle n_q \rangle^2 + \langle n_q \rangle$.  This hierarchy can be described by a diagonal ensemble with density matrix  
%
%*******************  diagonal density matrix  *********************
\begin{equation}
\rho_{\rm diag} =  e^{- \sum_{q > 0} \lambda_q n_q}   \delta_{n_q, n_{-q}} \ \ ,
\label{diag.eq}
\end{equation}
%****************************************************************
%
where $\lambda_q$ is an effective temperature chosen to fix $\expval{n_q}$ \cite{Dora:2012}.  This ensemble preserves  correlations  between modes with opposite momenta, $\langle n_q n_{-q} \rangle \equiv  \langle n_q n_q\rangle$, arising from interactions between them.  The density matrix in Eq.~(\ref{diag.eq})  is not equivalent to the generalized Gibbs ensemble density matrix \cite{Rigol:2007md,Polkovnikov:2011gg,Caux:2012cg,Ilievski:2015ap}; it is constructed from non-local degrees of freedom and does not include any information about non-particle conserving correlations \cite{Ilievski:2015ap}. 

We compute the bosonic entropy via $S_b = -2\Tr \rho_{\rm diag} \log \rho_{\rm diag}$, with the factor of two originating from the diagonal nature of the ensemble \cite{Gurarie:2013mu,Piroli:2017un}. Exploiting the fact that there are no interactions between the bosonic density waves, their entropy density is
\begin{align}
    \mathfrak{s}_b \equiv \frac{S_b}{N} &= {2\over N}  \sum_{q>0}  \left[\langle n_q\rangle \ln \langle n_q \rangle - (1+\langle n_q\rangle )\ln(1+\langle n_q\rangle) \right]\ . 
\label{eq:SbK}
\end{align}
Here, the sum runs over positive momenta only due to the constraint between modes in Eq.~(\ref{diag.eq}), enforcing the same occupancy of positive and negative $q$-modes and thus the modes with negative momenta do not contribute.

In the following, we argue that the post-quench entropy Eq.~(\ref{eq:SbK}) evaluated for the LL model is equal to the density of asymptotic spatial entanglement entropy for the underlying fermions.  This is motivated by the result that in equilibrium and at low temperatures, the bosonic entropy  is equal to the thermodynamic entropy of the microscopic interacting fermion system \cite{giamarchi:2004qu}.  For an integrable system after a quantum quench, the  Yang-Yang entropy computed from the occupation numbers of the exact eigenstates of the interacting system is  known to agree with the asymptotic steady state spatial entanglement entropy per particle \cite{Yang:1969yy,Alba:2015et,Alba:2017ph}, computed as the von Neumann entropy
\begin{equation}
    S(t;\ell) = -{\rm Tr} \left[ \rho_{\ell}(t) \ln \rho_{\ell}(t)\right] \ ,
\label{eq:SvN}
\end{equation}
where $\rho_\ell(t)$ is the post-quench spatially reduced density matrix.

We define a quadratic Luttinger model by demanding that it faithfully reproduces the first few oscillations of the static structure factor as a function of waiting time after the quench. We then compute the entropy of this LL model from its asymptotic diagonal ensemble density matrix. In this way, we approximate  the exact eigenstates of an interacting fermion model with those  of the Luttinger model, and propose  in analogy to the Yang-Yang entropy that Eq.~(\ref{eq:SbK}) computed from the LL mode occupancies is equal to the density of spatial entanglement.  For the LL model, $\langle n_q \rangle$ can be determined in an unambiguous manner from the structure factor by using Eq.~\eqref{eq:SqtSM}.  The above approximation seems reasonable as  both the microscopic and the LL model are similarly  constrained integrable models.  We
test this hypothesis with an extensive numerical experiment, and find excellent agreement between the post-quench diagonal ensemble entropy Eq.~\eqref{eq:SbK} and the asymptotic spatial entanglement entropy from Eq.~\eqref{eq:SvN}.

{\em Demonstration for a lattice model:} We consider an integrable one-dimensional model of spinless fermions described by
\begin{equation}
    H =   -J\sum_{i=1}^{L}\left(c^\dagger_{i} c^{\phantom{\dagger}}_{i+1} + h.c. \right) +  \theta(t) V \sum_{i=1}^{L} c^\dagger_i
    c_i c_{i+1}^\dagger c_{i+1}    
\label{eq:H-JV}
\end{equation}
where the first term corresponds to free fermions ($H_0$), and the second term to a nearest neighbor interaction $V$, switched on at time $t=0$ ($H_1$ with $\theta(t)$ the Heaviside step function). The equilibrium phases of the post-quench Hamiltonian at half-filling ($N=L/2$) are known via a mapping to the $XXZ$ spin-model \cite{DesCloizeaux:1966,Yang:1966yc}.  For $-2 < V/J < 2$ the ground state is  a quantum liquid, with a first order transition at $V/J=-2$ to a phase separated solid (ferromagnet in the spin language).  At $V/J=2$ there is a continuous transition to an insulating state with staggered density wave order (antiferromagnet in the $XXZ$ model).  The quantum liquid regime at low energies it is described by the linear hydrodynamics of the Luttinger model with interaction parameters $K$ and velocity $v$ given by
\begin{equation}
K = \frac{\pi}{2\cos^{-1}\left(-\frac{V}{2J}\right)};  \qquad 
\frac{v}{J} = {\pi} \frac{\sqrt{1-(V/2J)^2}}{\cos^{-1}(V/2J)}
\label{eq:vSM}
\end{equation}
where the lattice spacing has been set to unity. 

At $t=0$, the ground state $\ket{\Psi_0}$ of \Eqref{eq:H-JV} corresponds to non-interacting fermions, and after the sudden quench, the initial state evolves according to $\ket{\Psi(t)} = \mathrm{e}^{-i H t}\ket{\Psi_0}$ which is obtained via numerical exact diagonalization for systems sizes up to $L=26$ by exploiting translation, reflection, and particle-hole symmetries of the microscopic Hamiltonian. All post-quench observables are computed from $\ket{\Psi(t)}$ or its associated density matrix ${\rho} = \ket{\Psi(t)}\bra{\Psi(t)}$.  All code, data, and scripts necessary to reproduce the results of this work are included in an online repository \cite{repo}.

\begin{figure}[t]
\begin{center}
    \includegraphics[width=\columnwidth]{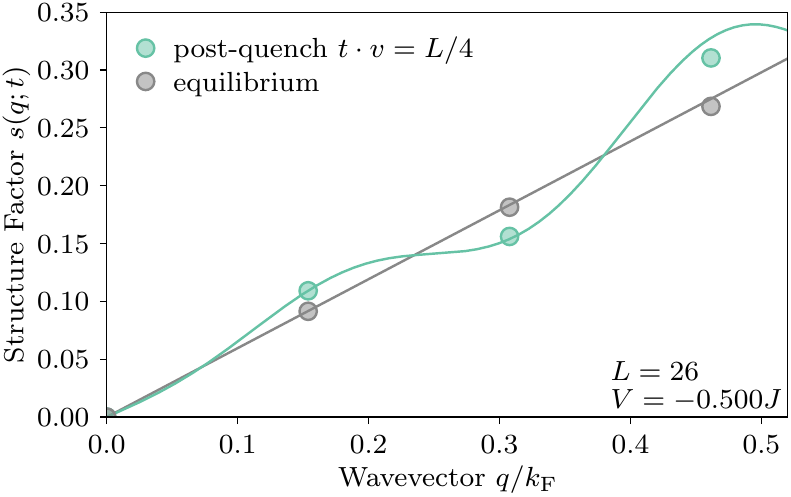}
\end{center}
\caption{{Comparison between equilibrium and post-quench structure factor.} The static structure factor at small wavevectors for a quantum quench to interaction strength $V=0.5J$ at fixed time $t \cdot v = L/4$ (corresponding to an extremum of the oscillations) deviates significantly from its equilibrium 
counterpart.  The solid lines correspond to Luttinger liquid predictions as defined in Eqs.~\eqref{eq:LLsq} and \eqref{eq:Sqtdef} using values of $K$ and $v$ computed from Eq.~\eqref{eq:vSM}.}
\label{fig:LL_structure_factor_comparison}
\end{figure}

In Fig.~\ref{fig:LL_structure_factor_comparison} we compare the post-quench static structure factor \Eqref{eq:Sqtdef} at a fixed time $t$ (corresponding to its first extremum for $q=2\pi/L$), with the time-independent structure factor of an associated equilibrium model having the same nearest-neighbor interaction strength, ($V = -0.5 J$ here) for $L=26$ sites at half-filling. 
%
% ------------------------------------------------------------------------------- 
% ------------------------------------------------------------------------------- 
%
In both cases, data points (circles) were obtained via exact diagonalization, and the theoretical Luttinger liquid predictions for small $q$ (solid lines) were computed from Eqs.~\eqref{eq:SqtSM} and \eqref{eq:LLsq}.  We have used the Bethe ansatz solution to convert the interaction strength to the effective parameters of the Luttinger model where $\expval{n_q} \equiv \expval{n_{q=0}} = \tfrac{1}{4}(K+K^{-1}-2)$, $\sinh 2\beta_q \equiv \tfrac{1}{2}(K-K^{-1})$ and $\omega_q = v q$ as discussed in the supplement \cite{supplement}.  At small wavevectors, the exact diagonalization results are in very good agreement with the Luttinger liquid predictions, both in equilibrium and for a quantum quench, while deviations start to increase as the wavevector approaches a finite fraction of $k_{\rm F}$.

\begin{figure}[t]
\begin{center}
\includegraphics[width=\columnwidth]{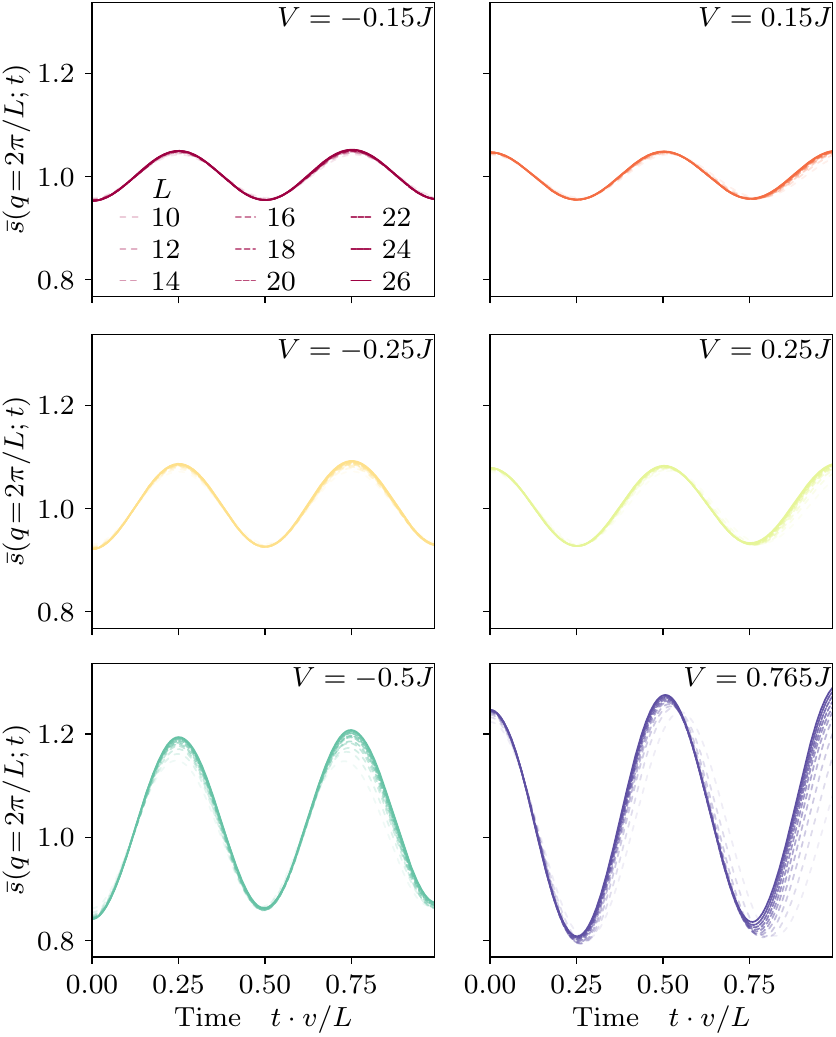}
\end{center}
\caption{Exact diagonalization results for the observation time dependence of the normalized static structure factor $\bar{s} = s(q;t)/s_{\rm eq}(q)$ after a quantum quench at the smallest value of momentum $q= 2 \pi /L$.  Panels corresponding to different final interaction strengths $V$. After a spline interpolation, the period of the oscillation is determined, allowing for determination of $\expval{n_q}, K$ and $v$ by taking into account the first one and a half periods. Time after the quench is measured in units of the velocity given in Eq.~\eqref{eq:vSM}.} 
\label{fig:LL_structure_factor}
\end{figure}

{\em Extracting the Bosonic Momenta Distribution}:
Exact diagonalization results for the post-quench time-dependence of the static structure factor (Eq.~\eqref{eq:Sqtdef}) $\bar{s}(q;t) \equiv {s(q;t)}/{s_{\rm eq}(q)}$ normalized by its equilibrium ground state value $s_{\rm eq}(q)$ are shown in Fig.~\ref{fig:LL_structure_factor} for a number of interaction strengths and system sizes for the smallest admitted wavevector, $q=2\pi/L$. In practice, the appearance of finite size effects are suppressed by rescaling the dimensionless time by the system size $L$, to achieve data collapse. 
%
% ------------------------------------------------------------------------------- 
% ------------------------------------------------------------------------------- 
%
The oscillatory structure seen in Fig.~\ref{fig:LL_structure_factor} predicted by the Luttinger liquid result $s_{LL}(q;t)/s_{LL}(q)$ in \Eqref{eq:SqtSM} can be exploited to recover the parameters of the underlying microscopic model as well as the $q$-dependent boson occupations $\expval{n_q}$ without the need for non-linear fitting.  For each interaction strength $V$, system size $L$, and wavevector $q$, we perform a cubic spline interpolation to the discrete time sampled $\bar{s}(q;t)$. The frequency of oscillations $\omega_q$ can then be independently determined from the location of the first three non-trivial extrema, corresponding to $1/2$, $1$ and $3/2$ periods respectively.  Integrating $\bar{s}(q;t)$ over these respective times yields:
\begin{equation}
    I_n(q) = \frac{\omega_q}{n \pi} \int_0^{n\pi / \omega_q} \dd{t} \qty[\bar{s}(q;t) + \delta \cdot t]
     = 1 + 2\expval{n_q} + \delta \frac{n\pi}{2\omega_q} 
\label{eq:In}
\end{equation}
where we have introduced a possible linear drift term $\delta \cdot t$ to confirm the consistency of the quadratic Luttinger liquid theory for density wave excitations. The drift term can be obtained from
\begin{equation}
    \frac{\pi \delta}{4\omega_q} = \qty(I_1 - I_{1/2}) = \qty(I_{3/2} - I_{1}) = \frac{1}{2}\qty(I_{3/2} - I_{1/2})
\label{eq:delta}
\end{equation}
and for all values of $q < k_{\rm F}$, we find $\abs{\delta}/J \le 2 \times 10^{-4}$, justifying the LL form of the static structure factor introduced in Eq.~\eqref{eq:SqtSM}. We thus ignore any short time drift in our subsequent analysis.  

Next, we can similarly determine the $q$-dependent boson occupations from the $I_n$ by combining:
\begin{equation}
    1 + 2\expval{n_q} = 2I_{1/2} - I_1 = \frac{1}{2}\qty(3I_{1/2} - I_{3/2}) = 3I_1 - 2I_{3/2}
\label{eq:nqfit}
\end{equation}
where in practice we determine $\expval{n_q}$ and its uncertainty as the average and standard error of the three different measurements.  Finally, the pre-factor of the oscillating term can be found from the extremal values at $t = 0, \pi/2\omega_q, \pi/\omega_q$. 

Thus, utilizing Eqs.~\eqref{eq:In}--\eqref{eq:nqfit} we extract $\expval{n_q}$, $\omega_q$ and $\sinh (2\beta_q)$ from our exact diagonalization data (a numerical experiment). Performing finite size scaling for each $q$ and extrapolating to $q\to 0$ the LL boson occupation number can be computed as shown in Fig.~\ref{fig:nLL} as a function of interaction strength. 
%
% ------------------------------------------------------------------------------- 
% ------------------------------------------------------------------------------- 
%
The success of this procedure can be independently confirmed by comparing the extracted values with those predicted from bosonization: $\expval{n_{q\to0}}  = (K + K^{-1} - 2)/4$ using Eq.~(\ref{eq:vSM}) which produces the solid line in Fig.~\ref{fig:nLL}.  As can be seen, the agreement is excellent over a wide range of interaction strengths suggesting that the Luttinger parameter $K$ could also be estimated in experiments using this procedure.  With access to $\expval{n_q}$ we can now directly employ Eq.~\eqref{eq:SbK} to obtain the bosonic entropy.

\begin{figure}[t]
\begin{center}
\includegraphics[width=\columnwidth]{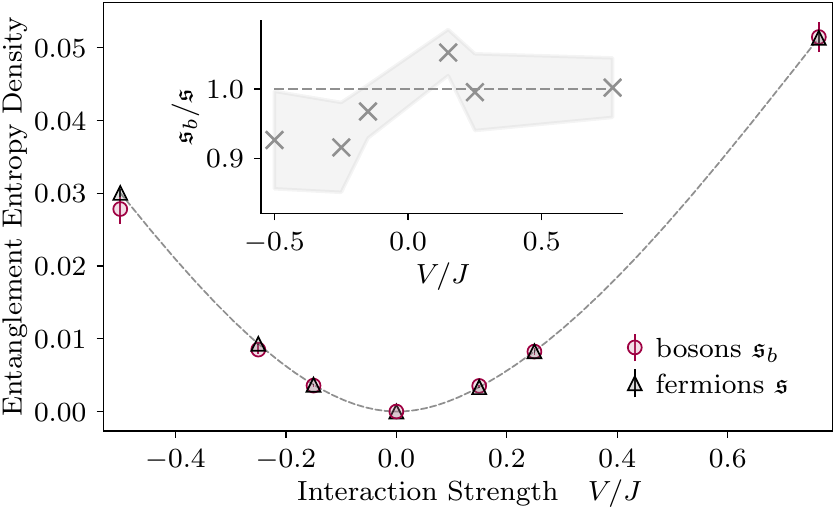}
\end{center}
\caption{Entanglement entropy density. We compare the steady state spatial entanglement entropy density $\mathfrak{s}$ for fermions with the entropy density $\mathfrak{s}_b$ computed from a bosonic diagonal ensemble of density excitations Eq.~(\ref{eq:SbK}). The bosonic occupation numbers $\langle n_q\rangle$ were obtained from an analysis of the data presented in Fig.~\ref{fig:LL_structure_factor}.  The dashed line is a guide to the eye. The ratio of the two entropy densities is consistent within error bars (inset).}
\label{fig:entropyComparison}
\end{figure}

{\em Comparison with spatial entanglement entropy:} This bosonic entropy can then be compared with the steady state ($t\to\infty$) value of the spatial entanglement entropy computed from Eq.~\eqref{eq:SvN}.   To this end, we determine the time dependence of the system's state after the quantum quench $\ket{\Psi(t)}$ via exact diagonalization and compute the reduced density matrix of a subsystem of macroscopic size $\ell=L/2$ by tracing over half spatial modes available to the fermions.  By applying the finite size scaling and temporal extrapolation procedure described in a previous work by the authors \cite{DelMaestro:2021ue} the density of the entanglement entropy  $\mathfrak{s} \equiv \lim_{L\to\infty}\lim_{t\to\infty}  S(t,L/2)/(N/2)$ 
can be obtained, where $N/2$ is the average number of particles in the spatial sub-region.  In Fig.~\ref{fig:entropyComparison} we show the comparison between the fermionic $\mathfrak{s}$ and bosonic $\mathfrak{s}_b$ entropy densities.
%
% ------------------------------------------------------------------------------- 
% ------------------------------------------------------------------------------- 
%
The agreement is within error bars estimated from uncertainties in the two finite size scaling procedures.  

\emph{Conclusions}:
In this paper, we have introduced a protocol to extract 
bosonic occupation numbers from the post-quench time evolution of the static structure factor in a fermionic lattice model.  Such a protocol could be implemented in current generation experiments of ultra-cold atomic gasses via access to the density-density correlation function after a sudden change in the system. The resulting occupation numbers of bosonic density waves can be used to compute their entropy within a diagonal ensemble of non-interacting bosons.  We compare this entropy with the exact steady state entanglement entropy of the microscopic fermionic model under a spatial bipartition and find excellent agreement between the two. This provides evidence that the fluctuating degrees of freedom of the approximate Luttinger model and the exact Bethe ansatz solution contribute equivalently to the steady state entropy density.  While the model considered here is integrable, the method can be extended to include integrability breaking perturbations.  In this case, any disagreement between the two entropies could indicate additional sources of entanglement generation and dynamics due to the reduction in conservation laws. 

\acknowledgements
This work was supported in part by the NSF under Grant No.~DMR-1553991 and 2041995 and the Deutsche Forschungsge-meinschaft (DFG) under Grants No. RO 2247/11-1 and No. 406116891 within the Research Training Group RTG 2522/1. A.D. expresses gratitude to the Institut f\"ur Theoretische Physik, Universit\"at Leipzig for hospitality during the initial phase of this work.

% ---------------------------------------------------------------------------------

% ---------------------------------------------------------------------------------
\FloatBarrier

\bibliography{refs}

\ifarXiv
    

\section*{Supplementary material}

# Supplemental Material for: "Measuring post-quench entanglement entropy through density correlations"

Adrian Del Maestro, Hatem Barghathi and Bernd Rosenow

In this supplement, we derive the time-dependent structure factor in the Luttinger liquid (LL) model after a quantum quench.

As bosonization captures the low energy and long distance physics of the model, the existence of discrete lattice sites is not important and we can adopt a continuum formulation. We represent the smooth part of the particle density as the spatial derivative of the displacement field $\rho(x) = (1/2\pi)\partial_x \phi(x)$, which in turn can be expanded in terms of normal modes as

$$\phi(x) = \sum_{q \neq 0} \sqrt{\frac{2\pi}{|q|L}} \left(a_q e^{iqx} + a_q^\dagger e^{-iqx}\right) , \tag{S1}$$

where the creation and annihilation operators satisfy the standard commutation relations $[a_q, a_{q'}^\dagger] = \delta_{q,q'}$. Then, the Fourier components of $\phi_q = \int_0^L dx \phi(x) e^{-iqx}$ are given by

$$\phi_q = \sqrt{\frac{2\pi L}{|q|}} \left(a_q + a_q^\dagger\right) . \tag{S2}$$

Since the Fourier transform of the nearest neighbor interaction of the fermionic Hamiltonian Eq. (8) is momentum dependent, we consider a finite range interaction in the following. For the moment we assume that the fermion dispersion can be considered to be linear over the range of momenta around the Fermi wave vector for which the Fourier transform of the interaction is nonzero, giving rise to the Hamiltonian

$$H = \sum_{q \neq 0} |q| \left[ v a_q^\dagger a_q + \theta(t) \frac{g(q)}{2} \left(2 a_q^\dagger a_q + a_{-q} a_q + a_q^\dagger a_{-q}^\dagger\right) \right] . \tag{S3}$$

For notational convenience, we now introduce the quantity $K_q = 1/\sqrt{1 + 2g(q)/v}$. After the quantum quench, the low energy sector of the Hamiltonian is that of a LL with parameter $K = K_{q=0}$. The equal time density-density correlation function at observation time $t$ after the quantum quench is given by

$$\langle \rho_{-q}(t) \rho_q(t) \rangle = \frac{|q|L}{2\pi} \left\langle \left[a_{-q}(t) + a_{-q}^\dagger(t)\right]\left[a_q(t) + a_q^\dagger(t)\right] \right\rangle . \tag{S4}$$

We now express the bosons $a_q$ in terms of the eigenstates $b_q$ of the interacting Hamiltonian via

$$a_q(t) = \cosh \beta_q b_q e^{-ivqt} - \sinh \beta_q b_{-q}^\dagger e^{ivqt}$$
$$a_{-q}^\dagger(t) = \cosh \beta_q b_q^\dagger e^{ivqt} - \sinh \beta_q b_q e^{-ivqt} \tag{S5}$$

with $\sinh^2 \beta_q = (K_q + K_q^{-1} - 2)/4$. Defining the bosonic occupation number

$$n_q = b_q^\dagger b_q , \tag{S6}$$

with $\langle n_q \rangle = \frac{1}{4}\left(K_q + \frac{1}{K_q} - 2\right)$, we find

$$\langle n_{-q}(t) n_q(t) \rangle = \frac{L K_q |q|}{2\pi} [1 + 2\langle n_q \rangle + \cos(2\omega_q t) \sinh(2\beta_q)] . \tag{S7}$$

Using the relation $s_{LL}(q;t) = (1/N) \langle n_{-q}(t) n_q(t) \rangle$, and the definition of the Fermi momentum $k_F = \pi N/L$, we obtain the result Eq. (3). In the presence of fermionic band curvature, the bosonic Hamiltonian corresponding to the model Eq. (8) contains nonlinear operators like $(\partial_x \phi(x))^3$, which lead to a coupling of bosonic modes, and a time dependence of mode occupancies $\langle n_q \rangle$. For this reason, we analyze the time dependence of the structure factor only for short times after the quantum quench, and allow for a possible time dependence by adding a linear in time drift term. From our numerical analysis, we find the time dependence of $\langle n_q \rangle$ to be very small.


\fi

\end{document}